\documentclass[manuscript]{acmart}

\usepackage{tabularx}
\usepackage{array}
\usepackage{pdflscape}

\AtBeginDocument{%
  \providecommand\BibTeX{{%
    \normalfont B\kern-0.5em{\scshape i\kern-0.25em b}\kern-0.8em\TeX}}}

\setcopyright{rightsretained}
\copyrightyear{2023}
\acmYear{2023}
\acmDOI{XXXXXXX.XXXXXXX}

\begin{document}

\title{Surveying a Landscape of Ethics-Focused Design Methods}

%\email{webmaster@marysville-ohio.com}
%\affiliation{%
%  \institution{Institute for Clarity in Documentation}
%  \streetaddress{P.O. Box 1212}

%\author{Lars Th{\o}rv{\"a}ld}
%\affiliation{%
  %\institution{The Th{\o}rv{\"a}ld Group}
 % \streetaddress{1 Th{\o}rv{\"a}ld Circle}
 % \city{Hekla}
 % \country{Iceland}}
%\email{larst@affiliation.org}

%\affiliation{%
 % \institution{Inria Paris-Rocquencourt}

\author{Shruthi Sai Chivukula}
\email{schivuku@iu.edu}
\affiliation{%
  \institution{Indiana University}
  \streetaddress{901 E 10th St Informatics West}
  \city{Bloomington}
  \state{Indiana}
  \country{USA}}

\author{Ziqing Li}
\email{li3242@purdue.edu}
\affiliation{%
  \institution{Purdue University}
  \streetaddress{401 N Grant Street}
  \city{West Lafayette}
  \state{Indiana}
  \postcode{47907}
  \country{USA}
}
\author{Anne C. Pivonka}
\email{apivonka@purdue.edu}
\affiliation{%
  \institution{Purdue University}
  \streetaddress{401 N Grant Street}
  \city{West Lafayette}
  \state{Indiana}
  \postcode{47907}
  \country{USA}
}
\author{Jingning Chen}
\email{chen3501@purdue.edu}
\affiliation{%
  \institution{Purdue University}
  \streetaddress{401 N Grant Street}
  \city{West Lafayette}
  \state{Indiana}
  \postcode{47907}
  \country{USA}
}
\author{Colin M. Gray}
\email{gray42@purdue.edu}
\affiliation{%
  \institution{Purdue University}
  \streetaddress{401 N Grant Street}
  \city{West Lafayette}
  \state{Indiana}
  \postcode{47907}
  \country{USA}
}

\renewcommand{\shortauthors}{Chivukula, et al.}

\begin{abstract}
Over the past decade, HCI researchers, design researchers, and practitioners have increasingly addressed ethics-focused issues through a range of theoretical, methodological and pragmatic contributions to the field. While many forms of design knowledge have been proposed and described, we focus explicitly on knowledge that has been codified as ``methods,'' which we define as structured supports for everyday work practices of designers. In this paper, we identify, analyze, and map a collection of 63 existing ethics-focused methods intentionally designed for ethical impact. Building on results of a content analysis of these methods, we contribute a descriptive record of how these methods operationalize ethics, their intended audience or context of use, their ``core'' or ``script,'' and the means by which these methods are formulated and codified. Building on these results, we provide an initial definition of ethics-focused methods, identifying potential opportunities for the development of future methods to support ethical design practice.
\end{abstract}




\maketitle

%Note to the review committee- \textbf{REMOVE BEFORE SUBMISSION}
%We would like to note the review trajectory of this paper, beginning in the year 2020 when we first submitted a version to CHI'21, followed by DIS'21; then, R \& R process in journals such as Design Studies'21 and TOCHI'22. We have received positive reviews from all the reviews about the seminal and novel work done in the paper. These multiple submissions and revisions helped us with a thorough revision of the paper to clarify the fundamental definition of methods (throughout the paper), strengthened and created a framework to classify methodologies from methods (as in Section 3.2.2) clear framing of the paper (added Section 2.3), and improving the contribution of the analysis beyond presenting a collection of ethics-focused design methods (iteratively worked on Sections 5.2 and 5.3).

{\color{red}\textbf{Draft: August 26, 2022}}

\section{Introduction}
There is a growing interest in socially responsible design, evidenced by the efforts of practitioners and third-sector organizations alike in building awareness and support for ethically-centered design practices \cite{darkpatterns-website,VSDbook,description,ethicalos}. These efforts have frequently built upon academic discourses such as ethics \cite{Shilton2018-ro,Gray2019-wa,ethicistasdesigner}, values \cite{valuesatplay,VSDbook,Friedman2003-og}, moral philosophy \cite{VSDbook,Bietti2020-rb}, and critically-oriented models of participation \cite{Shilton2018-ro,Keyes2019-fe,Bardzell2018-hy}.  While substantial efforts have been made to describe the value-centered or ethics-focused methods landscape from a scholarly and empirical perspective (e.g, Value Sensitive Design \cite{Friedman2008-iw,VSDbook}, Values at Play \cite{valuesatplay}, ethical standards or codes \cite{Gotterbarn2018-ol}, policies \cite{Goddard2017-gf,ethicscode}), a specific landscape of ethics-focused methods that are intended to pragmatically support the actions of designers and technologists in their everyday work is less well defined.

In prior work, design and HCI researchers have defined and engaged with methods as cognitive and pragmatic supports \cite{Stolterman2008-ty,Gray2016-lq,Harrison2006-dr}, a means of encouraging creative production \cite{Jones1992-gf}, an enabler of dialogue and communication during design activity \cite{Rittel1984-ga}, and a way of bridging multiple disciplinary ways of knowing to inform effective practice \cite{Hanington2003-bz}. However, design methods by themselves do not contain any action or inherently prescriptive or binding directives \cite{Gray2016-tt}, but rather are tools which enable design activity through the knowledge they contain \cite{Harrison2006-dr,Gray2016-ux}, under the control of the designer who activates this knowledge in situated and pragmatic ways to support their design activity \cite{Gray2016-ux,Lowgren1999-oy,Stolterman2008-ty,Stolterman2012-fr,Gray2022-na}.  Thus, we seek to investigate not only the means of supporting design practices in a broad sense, but also seek to describe the relationship between knowledge bound up in methods to the potential activation of that knowledge to create the potential for socially responsible design practices. The investigation of knowledge in design methods points towards questions, such as: How is a method structured and codified? What are the constituent elements of a method? and What is the language used to describe a method? In parallel, the analysis of methods through an ethics-focused lens reveals yet more questions, such as: How do methods enable designers to identify and act upon potential social impacts? Can methods guarantee ethical outcomes? What kinds of methods exist to engage and enable designers to identify and act upon potential social impacts? How can such methods be described for their prescription to designers to lead to ethically-sound outcomes? and How are ethical concerns inscribed into the language of methods? We do not seek to answer these questions in full, yet reveal this landscape of questions to demonstrate the potential broader impact of this work.%, are still unavailable for design researchers to present or describe a method, not just its role in design activity.

%This calls for understanding the ontology of a method by destructuring it to reveal the knowledge embedded, the study of which is scarce in design literature. The answers to questions such as:
In this paper, we identify, analyze, and describe a set of existing ethics-focused methods designed to support design research and practice for a range of audiences. Building on a content analysis of 63 collected methods, we describe how these methods operationalize ethics, are framed for particular audience(s), and are built to convey specific types of knowledge and sensitizing concepts. Across this collection of methods, we have deconstructed the language and specifications from the method source to identify the intended audience(s), format of guidance, interaction qualities, utilization of existing knowledge or concepts, implementation opportunities within design processes, the ``core'' or ``script'' of the method. %, and the ways in which the method builds upon or refers to existing ethical frameworks.
These aspects of ethics-focused methods aid us in characterizing a current landscape of ethical support for practitioners, elucidating opportunities for the adaptation of existing methods and the creation of future ethics-focused or value-centered methods for supporting design research and practice. %We have taken this opportunity of presenting the results of content analysis of a way to provide the ontology of methods and provide vocabulary to convey methods.

The contribution of this paper is three-fold: 1) We identify and present a collection of existing ethics-focused or value-centered methods in order to map the space of current ethical support for designers (as in Figure \ref{fig:mastertable1} and \ref{fig:mastertable2}); 2) We deconstruct the language and specifications of these methods to describe the framing used for the intended audience(s) and describe the means by which ethics is operationalized, facilitating more detailed inquiry into how methods are constructed and how they might further support ethical awareness and action (as in Section \ref{findings}); and 3) We identify opportunities, synergies, and gaps in ethics-focused methods, providing a roadmap for the creation and adaptation of methods that are resonant with the needs of practitioners (as in Sections \ref{d1}, \ref{d3},and \ref{definition}).

%[[We uncover and define methods as epistemology or ontology as a vehicle to look into designer practice---need to revisit after discussion is complete]].

\section{Background Work}

\subsection{Design Knowledge and Methods}
The notion of ``design knowledge’’ has been extensively researched in the design and HCI literature, broadly defining what constitutes design knowledge (e.g., patterns of reflection \cite{Cross1984-sw}, ontologies \cite{Willis2006-qb}), levels of instigation of design knowledge in design activity \cite{Kolaric2020-xq,Goodman2011-ak}, and different types of design knowledge \cite{Lowgren1999-oy,Lowgren2013-jt}. For the purpose of this paper, we explore ``methods'' as a particular form of design knowledge that enables ``the creation of design states'' \cite{Kolaric2020-xq} that support and advance a designer's capability \cite{Lowgren1999-oy}, building upon decades of interest in identifying key aspects of design cognition and the support of design work through methods (e.g., \cite{Jones1992-gf,Alexander1977-rq,Papanek1971-fv}). We draw on the definition given by Stolterman and colleagues \cite{Stolterman2008-ty} of design methods as ``tools, techniques, and approaches that support design activity in [a] way that is appreciated by practicing interaction designers,'' and Gray's  \cite{Gray2016-tt} definition that describes design methods as ``tool[s] that allow designers to support thinking, reflecting and acting upon design activities.'' Both definitions are consistent with historical framings of methods in the design studies literature, such as Cross' \cite{Cross1980-pp} description of methods as ``step-by-step, teach-able, learnable, repeatable, and communicable procedures to aid the designer in the course of designing.'' Within this framing, we wish to further describe how methods-focused knowledge allows researchers to better understand design practices, including the identification of areas where there is stronger and weaker support. Prior research on the use of methods by practitioners has shown evidence that methods are largely selected and used based on emergent aspects of the design context, where practitioners leverage knowledge enabled through the use of \textit{tools} either for thinking or generating artifacts \cite{Stolterman2008-ty}. In this sense, methods are primarily activated through a ``mindset'' rather than a precisely defined way to conduct design activity, and the performance of any given method or combination of methods is dependent on how a designer chooses to appropriate methods to support their design work \cite{Gray2016-ux}.

We view methods as a form of design knowledge that does not function alone, but is rather activated through the designer's activity and judgment, reflecting on the design knowledge contained within the methods \cite{Gray2016-tt,Gray2022-na,Goodman2011-ak,Stolterman2012-fr}. This knowledge can be abstracted further to describe the \textit{repertoire} of an individual designer, which includes both stores of existing design precedent \cite{Schon1990-by} and larger assemblages of tool knowledge that Gray et al. \cite{Gray2016-lq} have previously referred to as a designer's \textit{conceptual repertoire}. Thus, in building and elaborating the inherent structures of existing design methods, we are able to point towards a \textit{conceptual repertoire} that is implicit in both design knowledge and use. This notion of a conceptual repertoire also builds on previous work in the HCI and design communities that has interrogated both prescribed and performative accounts of design practices, including both the exploration and performance of methods by practitioners from Goodman and colleagues \cite{Goodman2011-ak,Goodman2013-bb} and Reeves \cite{Reeves2019-wn}, and the differentiation between codification and performance proposed by Gray \cite{Gray2016-tt}. %In particular, we focus on the notion of method ``cores'' \cite{Gray2016-ux} to describe an inscribed potential for design moves contained within methods, driven by a synergistic overlap of context, script (embedded instructions), and the lived experience of the designer.
In this paper, we specifically leverage the definitional work by Gray ~\cite{Gray2022-na,Gray2016-tt} in characterizing the knowledge contained within design methods, with our analytic focus for this study drawing only on their articulation of \textit{prescriptive} and \textit{presentation-oriented} stances towards methods. A prescriptive or codification-oriented stance ``reveals the extent to which procedural and descriptive knowledge is bound up in the method itself,'' while a presentation-oriented stance ``describes how the method is communicated, packaged, and disseminated, focusing on the ways in which methods are articulated to their anticipated audiences'' \cite{Gray2022-na}. We explicitly exclude accounts of methods that primarily leverage the performative stance, defined as ``aspects of a method that are revealed only as the method is used in a particular context by a designer, often with implicit connections to the codified form of the method '' \cite{Gray2022-na}, since this is a space already investigated by numerous HCI and STS scholars (e.g., \cite{Goodman2011-ak,valuelevers,Steen2015-mt,Reeves2019-wn}).

We also build upon prior work that has defined and curated a range of methods to support design activity, with such work having a stated goal of describing methods in ways that are simple enough for designers to adapt, apply and combine different methods in various ways \cite{curedale2012design}. L\"{o}wgren and Stolterman \cite{Lowgren1999-oy} have built upon this notion of method reuse, stating that methods should be accessible, flexible, and adaptable for designers to apply either independently or alongside the designer's current ``toolbox'' in different contexts. Further, Stolterman \cite{Stolterman2008-ho} has claimed that any knowledge introduced into design practice should bear a ``rationality resonance,'' whereby the content of methods should resonate with the complexity of practice. Building upon this framing of building resonance into design methods, multiple scholars have sought to create and curate design method collections, including: methods for creativity and innovation to improve the range of design production \cite{Biskjaer2010-tg,mose2017understanding}; a collection of UX evaluation methods \cite{vermeeren2010user}; a classification of methods, including traditional, adapted, innovative methods, and methods for interpretation and analysis \cite{Hanington2003-bz}; a popular collection of UX research and design methods for design students and practitioners titled \textit{Universal Methods of Design} \cite{umod}; a design kit for Human-Centered Design practice by IDEO \cite{ideo}; a collection of product design methods and approaches known as the \textit{Delft Design Guide} \cite{Van_Boeijen2014-vp}; an overview of strategies and methods for design innovation titled \textit{Design. Think. Make. Break. Repeat} \cite{Tomitsch2018-eb}; and a set of generative approaches for design research  \cite{Sanders2012-om}. While the list of curated collections of methods is already substantial, and still growing, we intend to build upon these collections with an explicit focus on methods that are designed to support \textit{ethically-centered practice}, building upon existing language to describe methods while also proposing new vocabulary to conceptualize, categorize, and propose links within and among methods.

\subsection{Supporting Ethical Design Practice}
HCI, Science and Technology Studies (STS), and design researchers have previously explored ethical practice across multiple dimensions, including: theoretical accounts \cite{Friedman2003-og,Shilton2018-ro}, methodological descriptions \cite{VSDbook, valuesatplay, Manders-Huits2009-nw}, identification of pragmatic and practice-led work \cite{Gray2019-wa,darkpatterns-website,Gray2016-ej,valuelevers,Steen2015-mt, Chivukula2020-oy}, and philosophical accounts \cite{Verbeek2005-eq, Dilnot2005-ah}. When focusing on prior research contributions relating to methodology, we have identified numerous frameworks that propose methodological means for designers to engage in value discovery and implementation. Common and well-known methodologies include Value Sensitive Design (VSD; \cite{Friedman2008-iw,VSDbook}) and Values at Play \cite{valuesatplay}. Other researchers have proposed strategies for designers or technologists to advocate for values in practice contexts, including organizationally-focused approaches such as Shilton's ``Values Levers'' \cite{valuelevers} or van Wynsberghe's ``Ethicist as Designer'' \cite{ethicistasdesigner}. It is claimed that these strategies and methodologies can ``open new conversations about social values and encourage consensus around those values as design criteria'' \cite{valuelevers}; identify new ways to expose and reflect upon designers' responsibility or attitudes towards value-based decisions \cite{darkpatterns2,assholedesignproperties}; propose suggestions for critical and reflective technical practice \cite{Agre1997-yk}; foreground tools for value comprehension in particular contexts \cite{VSDbook}; provide practitioners with ethical codes for computing work \cite{ethicscode}; frame policies for ethical responsibility for organizations \cite{Goddard2017-gf}; and offer requirements for ethics curriculum for computing and engineering education \cite{Gotterbarn1998-la,Hess2018-vg}. This range of prior work illustrates the efforts of the HCI, STS, and design communities towards identifying opportunities for supporting ethically-focused work practices. However, as an additional point of complication, portions of this prior work has been critiqued regarding its lack of resonance in authentic work settings, or due to the lack of adequate translation of these practices from academia to practice \cite{Shilton2018-ro,Manders-Huits2009-nw,Gray2019-wa}. Finally, while collections of methods have become commonplace in the last decade, with texts such as \textit{Universal Methods of Design}~\cite{umod} now in regular use by practitioners and educators, none of these collections appear to include even an implicit focus on the ethical content of methods. In this paper, we focus our efforts on surveying the landscape of ethics literature through the framing of design methods, with the goal of gathering and characterizing the existing landscape of ethics-focused and value-conscious design methods.

\subsection{Framing the Paper and Our Positionality} \label{ourframing}
Building on this background, we would like to provide more details regarding the positionality of the research team, which includes background work we have conducted in the space of design methods, to better frame the main contribution of the paper. We have engaged in studying ``design methods'' since 2014, including early work focusing on method performance and the development of design competence [citations removed for anonymous review]%\cite{Gray2014-fk,Gray2016-ux}.
Since 2016, we have leveraged both our definitional %\cite{Gray2016-tt,Gray2022-na}
and practice-focused research
\cite{Gray2016-tt,Gray2016-ux}
on design methods in building a collection of ethics-focused methods [citations removed for anonymous review].

Definitionally, in this paper we build on Gray's definitional work \cite{Gray2016-tt,Gray2022-na} to consider the vocabulary of design methods, their nature, and function in design activity. We consider prescription and performance of a method to be related but analytically separable, with key distinctions among: 1) the \textit{prescription} of a method as prescribed or advised in the method's published description about its use in a design activity; 2) the \textit{performance of a method in a design activity} as used by a designer based on their awareness of knowing, applying, and/or adapting an existing design method; and 3) \textit{performance of a method as mediated by an ecological setting} by an individual or group of designers tapping into concerns of resonance with organizational culture and interactions with other professional roles. Given that our goal for this paper was to begin building and describing a collection of ethics-focused methods, we focus only on Gray's \textit{prescriptive stance} (with limited analysis of Gray's \textit{presentation-oriented} stance), analyzing methods only as they are described in published material. To illustrate these limitations, imagine a metaphor of an amorphous box (a method) having some information (design knowledge) inside it; objectively, the box has some inherent properties (the prescription of a method) as designed by the ``box designer'' (a method designer; see \cite{Gray2022-kv} for more details on method designer processes). If a person (a designer) interacts with, compresses, or even tears the box apart, they start to interact with the information inside it depending on their situation, context, and intentionality; these inherent properties circumscribe, but do not completely define, the potential actions that the box might support (the performance of the method). In this paper,
we limit our inquiry to the identification of prescriptive elements that are used to describe and ``language'' a method \cite{Gray2022-na} in terms of their: 1) embedded \textit{codification} or prescription that describes the ``core'' of the method and the sensitizing concepts used by the method designer; and 2) the means of \textit{presenting} of the method described through its (in)accessible publication format, type of guidance, and tangible medium of the method to enable interaction with the method. With this understanding of the meaning, form, and function of a method, we only build our analysis and contribution based on the prescription of the methods in their current forms of the accessible method description. The synthesis of our findings based on this prescription provides a foundation for future research on the performance of methods in a design activity and particular ecological contexts, as detailed in Sections \ref{d1}, \ref{d3}, and \ref{definition}.

Our data collection and analysis over a two-year period was shaped both by our continuing inquiry into the ethical complexity of technology practices and our interests in identifying pragmatic supports to increase practitioners' awareness of ethical concerns and ability to act. As part of this overarching project, we created the notion of an ``ethics-focused method'' focusing only the prescription of these methods as comprising of any method that includes one or more sensitizing concepts or theoretical commitments that relate to values or ethics and in doing so, inscribes or operationalizes ethical concerns. %without focusing on the actual performance of these methods in a design context.
We further describe ``ethics-focused methods'' in Section \ref{definition} to illustrate how they differ from generic design methods. Our main intention is to \textit{begin} the process of building a collection of ethics-focused methods, recognizing that we cannot provide an exhaustive list of such methods\footnote{Indeed, across in our two-year analysis process, we have discovered new (or merely newly discovered) practitioner- and researcher-focused methods that have an ethical focus. Thus, we frame our collection of 63 methods as a foundation for more collection---similar to the 100 methods contained in \cite{umod} which have since been extended to 125 entries.}, while also identifying opportunities to build new ethics-focused methods and adapt other generic methods to include a more explicit ethics focus.

\section{Our Approach}
To map the landscape of existing ethics-focused methods, we collected a total of 63 methods and conducted a content analysis \cite{Hsieh2005-vz,Neuendorf2016-as} to describe the knowledge contained in these methods. This content analysis included the characterization of these methods on various levels and dimensions that will be detailed below as a part of our analysis approach. The research questions addressed through this paper are as follows:
\begin{enumerate}
    \item What design methods have an ethical focus, and how is ethics operationalized in these methods?
    \item Who are the intended audience(s) for these methods?
    \item How are these methods described?
\end{enumerate}

\subsection{Researcher Positionality and Rigor}
Our approach in using content analysis \cite{Neuendorf2016-as} involved a process of constant reflexivity and researcher alignment, given the complexity of the method and the ill-defined nature of our topic of interest. All researchers involved in this process have taken design and qualitative research methods coursework, and/or were involved in previous research projects that used content analysis or similar qualitative or critical analysis methods. Additionally, all researchers had prior experience engaging with conventional design methods through classroom projects and/or professional design work. These experiences enabled our research team to identify and characterize these methods, as we collectively brought knowledge of a broad spectrum of ethics-focused knowledge and design expertise. We reflexively engaged in open and axial coding as a key part of our content analysis process, where we first began to code for all potential aspects of method prescription (open coding) and later identified meaningful abstractions of these codes (axial coding) to form robust answers to our research questions. In this process, we employed strategies such as coder comments to track and build consensus, note taking and memoing to create robust coding schemes at every stage, peer debriefing of each others' codes to improve the rigor of the analysis process, and regular conversations with the research team to ensure alignment with generated coding schema at each stage. Additionally, given that our analysis of methods focused on the \textit{prescriptive stance} \cite{Gray2022-na}, we sought to identify appropriate axial codes based on the method prescription and not the potential application or performance of the method.

\subsection{Data Collection}
Through a series of structured web searches between January to November 2020 to locate ethics-focused methods, we collected a list of 89 methods/ tools/ approaches to begin our collection. The searches began by considering the Value Sensitive Design (VSD) methods \cite{VSDbook}, which enabled us to characterize the nature and purpose of ethics-focused methods, leading to our web searches on Google, Google Scholar, and the ACM Digital Library. The following keywords were used for the search queries: ``ethics focused methods,'' ``ethical tools in design,'' ``ethics methods,'' ``HCI ethics and values methods,'' and other related combinations of these terms. The methods we located were considered to be part of our initial collection if they had a clear ethical valence or had a stated intention to produce ethical or socially-responsible outputs. No specific year ranges were used as filters; nevertheless, we sought to identify as many methods fitting our criteria as possible within both traditional academic literature and from practitioner sources, given the lack of clear and consistent language to search using a more traditional ``systematic review'' approach. We more fully define our inclusion and exclusion criteria below. The precise scope of ``ethical outputs'' was not defined until the end of the analysis, as our goal was to identify methods that broadly had social or human considerations during the design process that related to an ethical valence. Methods we identified were published between 2008 and 2020, but we acknowledge that our search strategy may have missed methods published prior to these years, or different terms might have been used to describe such methods. Given the half-life of internet sources, our main aim was to capture the broadest range of methods presented in academic and practitioner sources. All methods were collected in a spreadsheet with descriptors such as the title of the method, published year, author names, and source files (documents or web links). The source files aided us in accessing the method’s description, which was our primary unit of analysis.

\subsubsection{Exclusion and Inclusion Criteria}
For the purpose of our analysis, we sought to include any design method that was created with the intent of supporting value-centered, ethically-focused, or socially responsible decision making practices, as indicated by the method description. We recognized that some methods functioned as \textit{methodologies}, and other methods contained multiple sub-methods; in these cases, we sought to identify the smallest method unit for analysis to increase precision. Through our reflexive data collection and analysis process, we also identified several exclusion criteria to narrow our focus. First, we excluded any methods that were computational, algorithmic, and UI-focused. For example, by computationally- or algorithmically-focused, we refer to toolkits such as those that offer a Python package to computationally test for biases, and algorithms that are intended to mitigate bias in datasets and models, such as the AI 360 Fairness Kit\cite{aifairness360} that was created to support software developer work, since recent work has analyzed such publicly available tools \cite{Morley2020-rt}. Second, we excluded UI-focused packages such as the IF Data Patterns Catalogue \cite{dp-ui-catalog}, which includes a set of interface choices suggested for handling user data, since the focus of these contributions was largely visual rather than methodological. Third, we excluded codes of ethics \cite{Gotterbarn2018-ol,Wolf2019-nm} and technology or legal policies \cite{Young2019-zb} as past work has evaluated the role of these codes in professional practice \cite{Lere2003-rk,Buwert2018-ik}, and the focus of these tools is generally on professional practice and not specific to design decision making. Fourth, we excluded methods intended to improve accessibility (e.g., recommendations for optimizing screen reading) and inclusivity (e.g., general broadening of participation in digital technologies), since this is already a well-defined area of technology practice and scholarship. The same criteria applies to the Participatory Design literature which is already internally very consistent and well studied. We add this as a delimitation of our work as the conversations about methods and practices within these communities are already established. We anticipate future work linking all these different design knowledge bases, but this is out of the scope for this paper. Finally, we excluded any entries that were not clearly expressed as a design method; as an example of the latter type, Stark \cite{Stark2019-vk} proposed a translational means of involving artists for ``work to produce a sense of defamiliarization and critical distance from contemporary digital technologies in their audiences''; this frame could be used to specify a future method, but is not currently articulated in a method-like form. Using these inclusion and exclusion criteria, we identified a list of 83 ethics-focused or value-centered methods, tools, approaches, conceptual vocabulary, methodologies, or frameworks. At this stage, we recognized that we had a heterogeneous collection of theoretical frameworks, concepts, methodologies, approaches, methods, which led to a further classification effort as described in the following section. We went through a reflexive process to define various potential classification approaches to define a final set of actionable methods. % which was a complicated process.

\begin{table*}
\centering
\caption{Classification of Collected Ethics-Focused or Value-Centered methods}
\label{types}
\begin{tabularx}{\textwidth}{p{.35\textwidth}X}
\toprule
\textbf{Types of Framing} & \textbf{Examples} \\
\midrule
\textbf{Methods} & Detailed and described in Figure~\ref{fig:mastertable1} and \ref{fig:mastertable2}. \vspace{5px} \\
\textbf{Methodology} & Value-Sensitive Design \cite{VSDbook}, Values at Play \cite{valuesatplay}, and Research through Design Fiction \cite{rtd-fiction}. \\

\textbf{Theoretical Commitments} & Feminist HCI \cite{feministHCI,feministhci-parenthetical}, dark patterns \cite{darkpatterns1,darkpatterns2,darkpatterns3,darkpatterns4}, Data Feminism \cite{datafeminism} and others \cite{assholedesignproperties,chronodesign,designjustice,manifesto,inactionethics,7sins,responsibleai}. \vspace{5px} \\

\textbf{Conceptual Frames} & Speculative Design \cite{speculativeeverything}, Critical Design \cite{criticaldesign}, Reflective Design \cite{reflectivedesign}, and others \cite{adversarialdesign,postcolonialcomputing,queering}.\vspace{5px} \\
\bottomrule
\end{tabularx}
\end{table*}

\subsubsection{Classification of Collected Artifacts}
Building on our collection of 83 artifacts, we sorted them into four main categories based on their potential function in design activity. As shown listed in Table~\ref{types}, these functions include: components of methods identified through prescriptive and presentation-oriented stances, theoretical commitments, methodologies, and conceptual frames. We present these as non-exclusive framings to highlight that a selected artifact can fall under one or more of these framings.

\textbf{Methods:} These provide guidance on a practical level, indicating to the designer how they might apply, operationalize, or activate ethics and values in technology design work. For this paper, we identified a list of 63 ethics-focused or value-centered methods prescribed for design action, which will be referred to simply as ``methods'' throughout the remainder of the paper. We will further elaborate how these methods serve as the main contribution of this paper, and we focus on this set to answer our research questions.

\textbf{Theoretical Commitments:} These provide guidance to designers on a theoretical level by characterizing the designer's ethical commitments (e.g., Data Feminism \cite{datafeminism} and Ethical by Design: A Manifesto \cite{manifesto}), listing qualities required for building ethical outcomes (e.g., Feminist Interaction Design Qualities \cite{feministHCI} and dark patterns \cite{darkpatterns1}), describing existing designs that are manipulative or value-centered (e.g., Asshole designer properties \cite{assholedesignproperties} and Nodder's Seven Sins \cite{7sins}), or suggesting organizational structural changes to include ethicists to incorporate ethical reflection into the product (e.g., Ethicist as Designer \cite{ethicistasdesigner}). Theoretical commitments do not tell the designer precisely how to engage in some of these practices or point towards actionable ways of implementing the concepts; these commitments are not yet procedural in form or defined for the designers in a way that directly activates their principles in concrete contexts, but rather suggests the required perspectives and language that might be considered when creating a method. For example, Bardzell's Feminist HCI commitment \cite{feministHCI} lists qualities that ``characterizes feminist interaction'' such as pluralism, participation, advocacy, ecology, embodiment, and self-disclosure; these qualities could be used to create one or more methods for \textit{Feminist Interaction Design} by defining steps or other means by which designers could apply these qualities in their design work. Other examples that fall under theoretical commitments include dark patterns strategies \cite{darkpatterns2,darkpatterns3,darkpatterns4,darkpatterns-website}, Design Justice \cite{designjustice}, and In-Action Ethics \cite{inactionethics}.

Between prescriptive methods and theoretical commitments lie \textbf{methodologies} which include a theoretical framing or umbrella of relevant and appropriate practices that can be applied in a design situation, often without the suggestion of specific tools and techniques. As one example of a methodology, Values at Play \cite{valuesatplay} suggests that the user ``discover, analyze, and integrate values'' specifically to game design; however, this methodology can be applied across any design situation within these stages. Other examples that are included as methodologies in our collection include %Value Levers \cite{valuelevers},
Research through Design Fiction \cite{rtd-fiction}, and VSD \cite{VSDbook}. As these framings are non-exclusive, VSD falls under the framings of a methodology when considering approaches to engaging values in a design process, a theoretical commitment towards human values in design activity, a conceptual frame for ethics-focused knowledge in engineering design, and over time, has also resulted in 10 specific methods that leverage the VSD methodology. We also identified some artifacts that do not fall neatly under these categories but still appear to have ethics-focused qualities. For example, Values Levers \cite{valuelevers}, which is well cited in the STS literature, is potentially ``method-like'' in that it references opportunities for alignment or disruption in a design setting, however with no particular inputs and outputs. To not confuse these pragmatic framings with what we categorized as ``methods,'' we have categorized them as ``other'' to avoid confusion and exclude from our analysis.

\textbf{Conceptual Frames:} These provide guidance at an \textit{epistemological} level, providing a more expansive set of proposed practices and knowledge which point towards broader approaches to building knowledge. For instance, a Critical Design approach \cite{criticaldesign} focuses on non-affirmative design practices, recognizing the knowledge that is built in the process of creating design artifacts. Other examples that fall under conceptual frames include Adversarial design \cite{adversarialdesign}, Agonistic design \cite{Bjorgvinsson2012-jp}, Critical Design \cite{criticaldesign}, Postcolonial computing \cite{postcolonialcomputing}, Reflective Design \cite{reflectivedesign}, Speculative design \cite{speculativeeverything}, and the Queering of HCI \cite{queering}.

\begin{table*}
\centering
\caption{Codebook of method characteristics in a prescriptive and presentation-oriented stance.}
\label{Codebook}
\begin{tabularx}{\textwidth}{p{.25\textwidth}X}
\toprule
\textbf{Characteristic} & \textbf{Description} \\
\midrule
\textbf{Core} & The central mechanic of using this method that remains relatively stable during adaptation and use. Axial codes include: posture types (\textit{eliciting values, critically engaging, defamiliarizing}) and actions (\textit{consensus building, evaluating, framing, generating}).  \vspace{5px} \\

\textbf{Activated Ethics Framework(s)} & Ethics theor(ies) mapped as being activated through a method as a part of the ``outcome expected.'' Axial codes include: \textit{deontological, consequentialist, virtue, pragmatist, and care ethics}. \vspace{5px} \\

\midrule
\textbf{Primary Audience} & Intended users of the method. \\
\textbf{Discipline/Domain} & The framing discipline to which the method applies or through which it is framed. \\

\textbf{Published Format} & Dissemination strategy of the method and the accessibility to the method to the intended audience\\

\textbf{Context of Use} & Environmental or logistical aspects of using the method. Axial codes include: group types (\textit{team, individual}) and ecology types (\textit{industry,instructional context}). \vspace{5px} \\

\midrule
\textbf{Input} & Elements the method operates on, which is inputted by the user(s) of the method. Axial codes include: \textit{user information, design artifacts/ services, users/stakeholders, values, framing constraints, problem frame, scenarios/context, and research material. } \\

\textbf{Mechanics} & Actions expected from the users while using this method. Axial codes include: \textit{altering, storytelling, filtering, creating, mapping, and evaluating.}  \\

\textbf{Output} &  Tangible results after using this method. Axial codes include: \textit{values, concepts, research outcomes, evaluation results, users/stakeholders, opportunities, procedural information, and research outcomes.} \\

\textbf{Outcome Expected} & Expectations from the user(s) and the ways in which the output might be manipulated by the user(s).   \\

\textbf{Existing Method(s) used} & Established design methods/ methodologies that are referenced or used as part of using or building the method.  \\

\textbf{Design Process Steps} & Design phase in which the method is prescribed to be used or can be used. Axial codes include: \textit{a priori} phases from Universal Methods of Design~\cite{umod}.  \\

\midrule

\textbf{Type of Guidance} & Ontological description and knowledge proposed as a part of the method’s description. Axial codes include: \textit{steps, guidelines, framework, lens/ perspective, reflective questions, examples, heuristics, %epistemology, methodology,
and case study.}\vspace{5px} \\

\textbf{Primary Medium} & Tangible form in which the method has to be used or structured. Axial codes include: \textit{worksheet, template, cards, document/ guidebook, physical manipulatives, videos, idea/ practice, and game}.
\vspace{5px} \\

\midrule

\textbf{Sensitizing Concepts} & Established theoretical concepts that are used in this method and the theory that has given the method’s vocabulary. \vspace{5px} \\

\bottomrule
\end{tabularx}
\end{table*}

\subsection{Data Analysis}
Using the method descriptions as our unit of analysis, we describe our data analysis procedures in three broad steps, as guided by Neuendorf's content analysis approach \cite{Neuendorf2016-as}: 1) familiarizing ourselves with the data set; 2) creating and validating the coding scheme; and 3) performing open and axial coding.

\subsubsection{Familiarizing with Data}
We began our analysis process by performing a close reading of several methods to familiarize ourselves with the organization of knowledge and language used to characterize each method. Two researchers individually identified preliminary codes for nine methods (including a diversity of topics, audience, and goals), including potential descriptors that aided our team in characterizing the content of the methods, pointing towards an initial coding scheme. The focus of this content analysis was to describe the method and analyze its characteristics based on a clear reading of its text, and not the instigation, creation or evaluation of the method in the context of practice.

As a part of the initial coding scheme, the researchers listed candidate descriptors that ranged from the form of the method, its potential application in design processes, expected outcomes, intended audience, attributes, and means of interaction with the method. After multiple rounds of iteration and discussion among the research team, a preliminary codebook of descriptors was created as detailed in Table~\ref{Codebook}. The more robust axial codes underneath the broader descriptors were determined later in the analysis process. During this stage, we also began to identify open codes \cite{Braun2006-cd,Saldana2015-ey} to describe potential methods ``cores,'' alongside researcher-inferred assumptions about the potential primary audiences that may use the method. By ``core'' of the method, we refer to the central mechanic of using this method that remains relatively stable during adaptation and use \cite{Gray2016-ux,Gray2016-tt}.

\subsubsection{Creating and Validating Coding Schemes}
During the second stage of analysis, we focused on validating the overarching descriptors from the initial coding scheme by revisiting the same nine methods coded in the previous round. We used a linked set of spreadsheets to conduct and document the content analysis of all the collected methods. This approach (facilitated by the tool, AirTable\footnote{https://airtable.com}) aided us in clearly building the audit trail of our coding process and relating these codes to previous coding work, increasing the validity and robustness of our codebook. At this stage, we began with an iterative process of open and axial coding under each of the main descriptors as described in our final stage of analysis. For each descriptor, we reflexively moved through stages of open coding, identification of preliminary definitions in a codebook, and extended conversation among members of the research team. Through deliberation over multiple weeks and rounds of coding and revisions to the codebook, we identified a final codebook for each descriptor set. All codebook elements, and the use of these elements in the coding process, were evaluated by pair coding and all application of descriptors was discussed until full agreement was reached.

\subsubsection{Open and Axial Coding Descriptors}
With the high-level descriptors (Table~\ref{Codebook}) finalized and the researchers aligned in their understanding, we conducted open coding of the full set of methods using the codebook. Once this initial coding was completed, we used these open codes to identify axial codes within each descriptor, using the process described above. The final round of analysis included summative, top-down coding using the final descriptors and sets of axial codes, including the type of guidance, primary medium, input, mechanics, output, core; all axial codes are listed in the description column of Table~\ref{Codebook} in italics. The role of axial coding varied for each descriptor, and is detailed in the findings section below.

\section{Findings} \label{findings}
\begin{figure*}
    \centering
    \includegraphics[width=\textwidth]{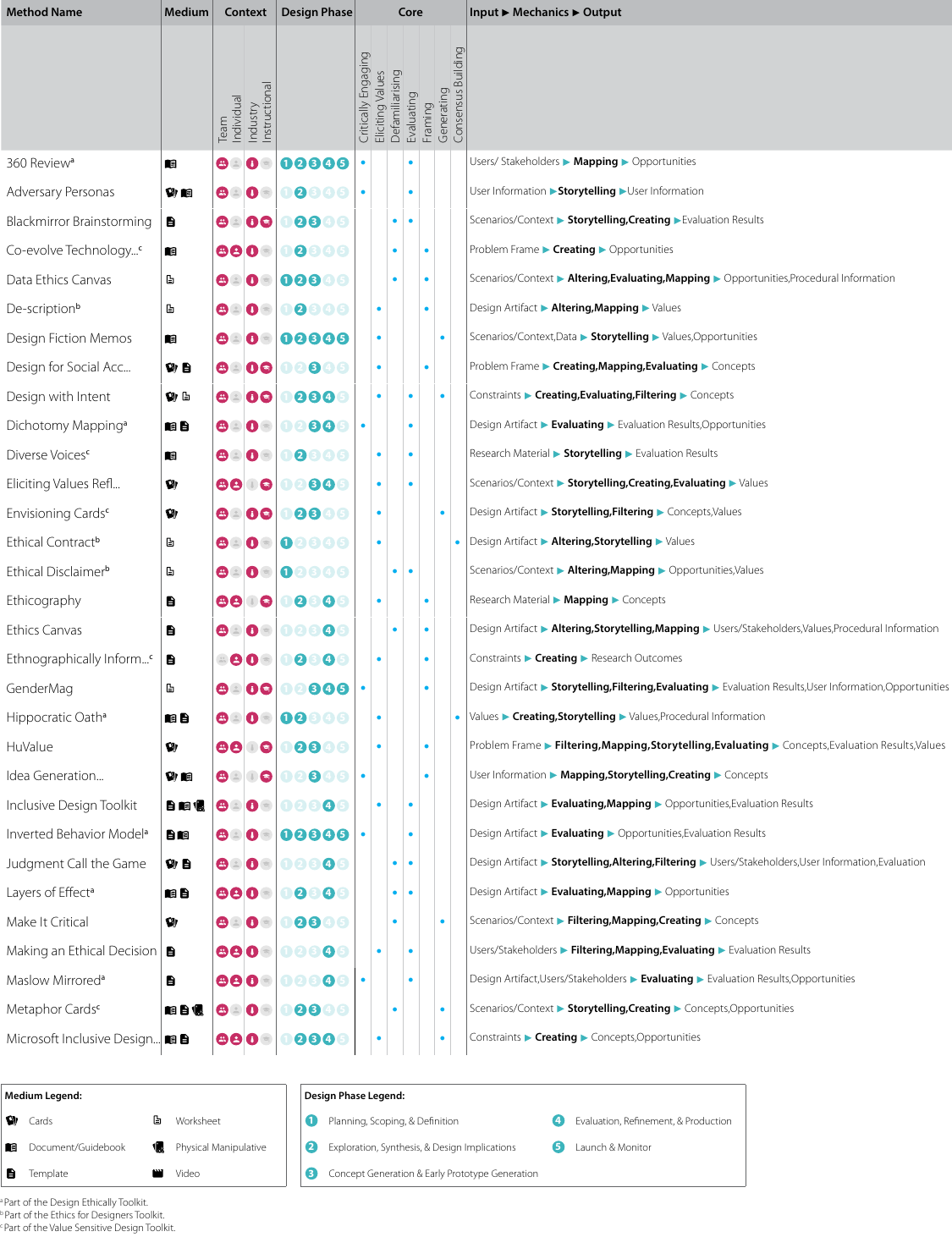}
    \caption{Methods (part 1) classified and organized by medium, context, design phase, core, input, mechanic(s), and output(s).}
    \label{fig:mastertable1}
\end{figure*}

\begin{figure*}
    \centering
    \includegraphics[width=\textwidth]{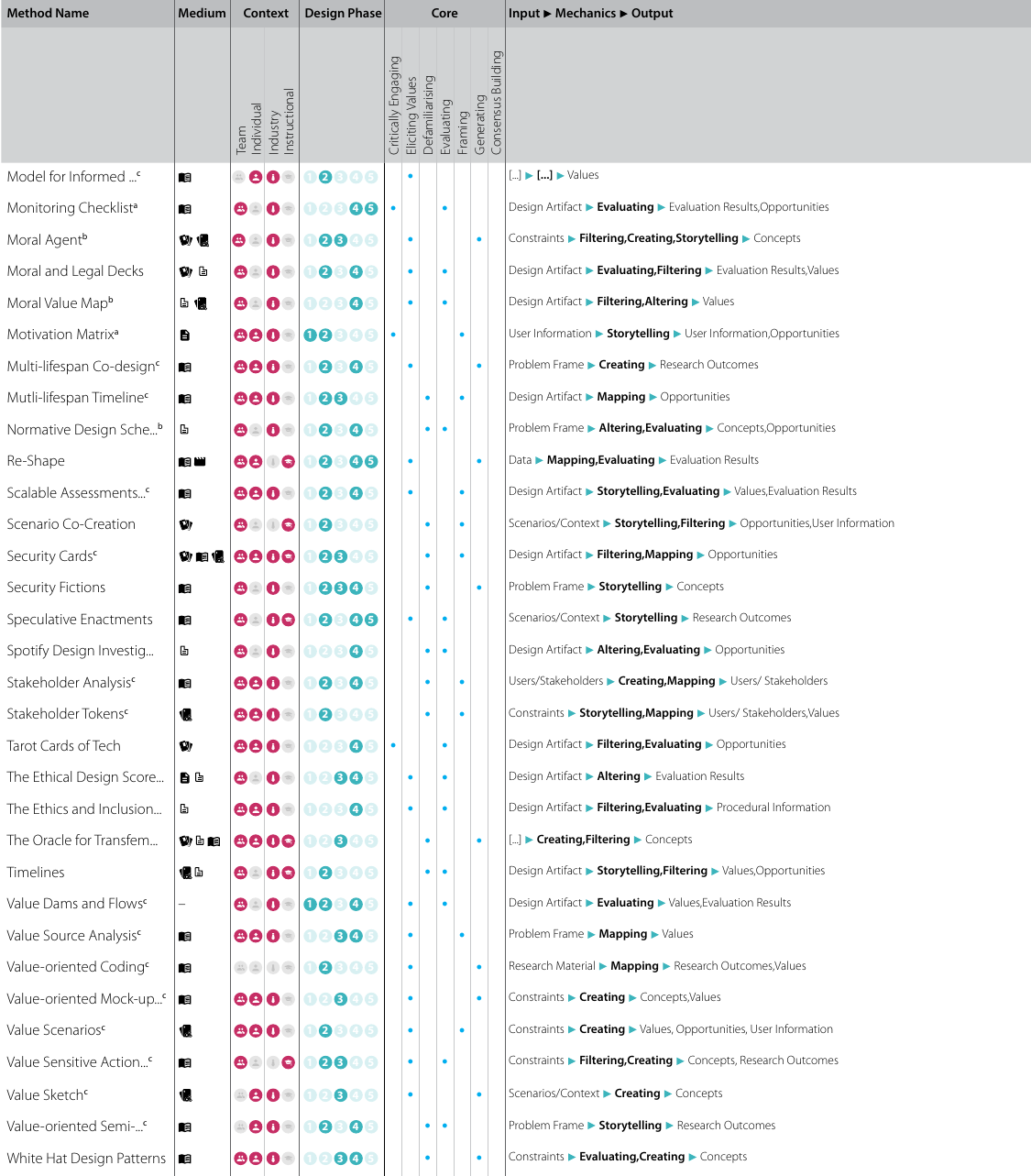}
    \caption{Methods (part 2) classified and organized by medium, context, design phase, core, input, mechanic(s), and output(s).}
    \label{fig:mastertable2}
\end{figure*}

In this section, we report on the findings of our content analysis, divided by research question. The three main subsections include: 1) The method's operationalization of ethics, where we describe the \textit{core} of the methods that hint towards ethical valence activated in the method; %and \textit{ethical frameworks} that are activated;
2) The intended audience for these methods; and 3) The formulation, articulation, and conceptual language used to describe these ethics-focused methods. A summary of the method descriptors is provided in Figures~\ref{fig:mastertable1}~and~\ref{fig:mastertable2}.

\subsection{RQ\#1: Operationalization of Ethics}
In this subsection, we identify how ethics or values were operationalized in these methods. We describe this operationalization through a property called the \textit{core} of the method; and 2) \textit{the ethical framework(s)} mapped to be activated in the method.

\subsubsection{Method Core}
By method \textit{core}, we refer to what we inferred as the central concept or essence of the mechanics of the method. Each method's core was identified from two groups: 1) \textit{Postures}: eliciting values (n=32), critically engaging (n=10), defamiliarising (n=21), and 2) \textit{Actions}: consensus building (n=2), evaluating (n=26), framing (n=20) and generating (n=15). We propose ``postures'' to be very specific to ethics-focused or value-centered methods, whereas the ``actions'' apply across any design method. Cores involving \textit{postures} target attitudes towards a certain action, either to identify an existing or generated list of values as a conceptual frame \textit{(eliciting values)}, engage in critical perspectives or theories drawn from critical theory as an epistemological argument \textit{(critically engaging)}, or take part in alternative forms of looking at existing concepts or forms of thinking \textit{(defamiliarising)}. Cores involving \textit{actions} encourage users to align their decision making with other stakeholders \textit{(consensus building)}, assess and validate the decision \textit{(evaluating)}, map the design space for using the method \textit{(framing)}, and produce design artifacts \textit{(generating)}. For example, the Inclusive Design Toolkit \cite{inclusivedesigntoolkit} had a core of ``eliciting values,'' since it proposes to evaluate a design artifact using values of inclusivity or accessibility. In Judgment Call the Game~\cite{judgementcallthegame}, the method core focuses on ``defamiliarizing,'' through which designers can ``evaluate'' a design scenario through reviews and ratings from the perspective of alternative users in the situation. Another example is Security Fictions~\cite{securityfictions}, which asks users of the method to ``defamiliarize'' themselves to think differently about security issues as they ``generate'' concepts to solve security threats. As these examples illustrate, the two groups of cores represent how ethics is operationalized based on the postures leading to those actions.

\subsubsection{Activated Ethics Frameworks}
We also identified how each method related to established ethical frameworks as a way to illustrate and provide us a vocabulary for how ethics were operationalized; drawing from multiple key texts in the philosophy literature, including Becker \cite{Becker2001-vt}, Kant \cite{Kant1785-is}, Aristotle \cite{Aristotle2000-hm}, and Gert \cite{Gert1984-nm}. We found evidence of multiple common framings or paradigms of ethics, including deontological, consequentialist, virtue, pragmatist, and care ethics. However, exclusive mappings of ethical paradigms to methods were problematic to identify due little indication by method designers which paradigm(s) drove the method, and by complications in how methods knowledge is prescribed more generally. Therefore, we present only a \textit{high level summary} of how we observed ethical paradigms to be intertwined in the methods we evaluated and do not insist on specific mappings of one or more ethical paradigms that align with each of the methods in our collection. As one example of the complications of identifying inscribed ethical frameworks through only a prescriptive stance, we might identify any ethics-focused method we analyzed to contain some deontological residue, even if the goal of the method was not to present a normative framing of practices that are always right or wrong. One explicit example that exemplified the ``duty ethics'' of the deontological paradigm is \textit{Ethical Contract}~\cite{ethicalcontract}, which requests that the users divide the ethical responsibilities in the project planning stage and physically sign the document (provided by the method) as a means of foregrounding their duty in accepting ethical responsibility. This method also includes commitments to virtue ethics by indicating a focus on the designer themself and their character. As another rare case of engagement with ethical paradigms in the prescription itself, the \textit{Normative Design Scheme}~\cite{normativedesignscheme} included spaces on a worksheet to assess design goals based on explicit references to virtue, consequentialist, and deontological ethics. Most methods could also be read in ways that foregrounded a consequentialist ethical paradigm, focusing on evaluating or considering consequences of design decisions, since all design efforts by definition result in an intentional reshaping of the world. A strong example of a consequentialist method includes \textit{Black Mirror Brainstorming}~\cite{blackmirror-brainstorming}, which encourages the designer to use critically-focused brainstorming to identify potential negative future impacts of designed outcomes. More rarely, an ethical paradigm was used to motivate specific kinds of guidance as a driving theoretical commitment rather than an explicit ethical frame, as with care ethics in Re-shape~\cite{reshape-paper}. Finally, since we approach methods through multiple stances that include acknowledgment of codification and performance orientation, all methods could also point towards the pragmatist ethical paradigm in some sense, because all methods are intended to support normatively complex and contextually-grounded ethical decision-making by designers. \textit{Moral Value Map}~\cite{moralvaluemap} represents a method that explicitly foregrounded this pragmatic dimension of ethical engagement, where the designer to choose relevant human values relevant to ``your design,'' but virtually all methods we evaluated contained the potential for such pragmatic engagement depending on how prescriptive elements are read, interpreted, and used to frame the potential for action.

\subsection{RQ\#2: Intended Audience}
In this subsection, we describe the implied audience and context for these methods across the following descriptors: 1) the \textit{primary intended audience} of the methods; 2) the defined \textit{context of use} of these methods; and 3) the \textit{published format} as a means of disseminating the methods.

\textbf{Primary audience.} We relied upon the method developer's identification of the intended audience from the method description to infer the \textit{primary audience} of the method. The stated audience types included: educators (n=3), academic researchers (n=15), students (n=6), industry practitioners (n=54), or anyone (n=1). Of the 63 methods, 13 did not explicitly state the primary audience (leaving the researchers to infer the audience), while the remaining methods mentioned their intended audience in the method description. For methods that intended to encourage conversations and collaborations among industry practitioners, sub-audiences primarily targeted design professionals (n=28), technology professionals (n=17), industry researchers (n=12), policy makers (n=2), and managers (n=1). The majority of methods had an intended audience from only one category rather a combination of multiple stakeholders.

\textbf{Context of use.} We describe the \textit{context} in which the method is intended to be used within two categories. First, we sought to identify whether a \textit{team and/or individual} was the ideal group size for the method to be used. Methods that appeared to be designed for a group as the intended audience were the most common (n=35), while individuals were a minority (n=4); methods identified for use by either a group or an individual were coded under both categories (n=23). Second, we sought to describe whether \textit{industry and/or instructional} settings were the primary ecological setting for methods to be used. Methods were coded to represent the anticipated ecological setting, including industry work for practitioners (n=45), an instructional setting for students and educators (n=3), or research in an academic context (n=4). A minority of methods (n=9) anticipated use both in industry and academia.

\textbf{Published format.} We identified the \textit{published format} for each method to identify its dissemination strategy or availability, thus revealing assumptions regarding the intended audience or the type of knowledge building the method represented. The publication formats we identified included academic papers (n=28), websites (n=20), blog posts (n=16), books (n=4), and unpublished (n=1; \cite{makeitcritical}). It is interesting to note that the majority of the methods we analyzed were published as academic papers, while the primary intended audience of these methods were industry practitioners. The limited availability of these papers behind a paywall perhaps brings into question the accessibility of these methods for the intended audience.

\subsection{RQ\#3: Formulation, Articulation, and Language}
In this subsection, we describe and characterize the collection of methods in three ways: 1) \textit{Formulation} of the methods as scripted to define input-mechanics-output, existing frameworks/structure and design process implementation of these methods; 2) \textit{Articulation} of these methods to the audience to describe type of guidance and medium of these methods; and 3) \textit{Language} that formed core of these methods. This section answers our research question \#3.
\subsubsection{Formulation of Methods}
We coded the ``script'' of these methods which illustrate the structure and interaction with these methods. We describe the formulation of the methods through major categories: 1) \textit{input} required, \textit{mechanics} of interaction and \textit{output} generated from the method; 2) \textit{existing frameworks or methods} used to build the method; and 3) practical implementation of the method in a \textit{design process}. We present different axial codes, definitions and examples in the paragraphs below.

\textbf{Input--Mechanics--Output.} The input--mechanics--output sequence describes how the methods are formulated, pointing towards potential patterns of performance. We identified ten salient \textit{inputs}, six action-oriented \textit{mechanics}, and eight tangible \textit{outputs} across the collection of methods. We will describe each element of the sequence separately below using a variety of methods as supporting examples. We have observed that despite a similar input, the change in mechanic has the potential to result in different outputs, giving us an opportunity to explore the interactions among these three elements. We provide more details regarding the patterns of interaction in the discussion section.

\textit{ \textbf{Input.}} We coded the required materials or knowledge the method developer wants the users to \textit{input} as a means to proceed with the method. The identified inputs include design artifact/service/business (n=23), research material (n=3), problem frame (n=8), constraints (n=9), users/stakeholders (n=4), user information (n=3), scenarios/context (n=10), values (n=1), data (n=2), and one method without any required input. Each method may include a wide range of inputs, but we chose to exclusively code only the most salient input required for each method to provide a more precise set of entry points. Methods with the input of \textit{design artifact/service/business} require the user to select an existing design material, product, or business service. For example, Multi Lifespan Timeline \cite{multilifespand-timeline-co-design} requires the user to provide a technological design artifact as an input in order to map how this artifact would exist in a social context at different timelines beyond the product lifecycle.  Methods with \textit{research materials} as an input require users to bring materials constructed for research purposes, such as interview protocols, co-design materials, tech policy documents, and others. For example, Scenario Co-Creation Cards \cite{scenarioco-creationcards} require an interview protocol along with the cards for conducting value-eliciting interviews with a culturally-diverse population. Other examples in this category include the Value Sensitive Action-Reflection Model \cite{actionreflectionmodel}, which requires the input of co-design materials, and Diverse Voices \cite{diversevoices}, which requires a tech policy document for expert panel discussions. Some methods required users to input their \textit{problem frame} in order to construct, define, and approach a design space through the lens of that method. We differentiate this input from \textit{constraints} based on their level of definition; whereas constraints are expected to be precise, problem frames are frequently more open-ended. \textit{Constraints} include stakeholder requirements, a project brief, time constraints, limited resources, or other explicit requirements that must be met. Methods that require \textit{users/stakeholders} or \textit{user information} as inputs encourage the description of stakeholder needs in relation to method-guided decision making. \textit{Users/stakeholders} indicates who the design is created or evaluated on behalf of, or who needs to be considered in decision making, whereas \textit{user information} describes user needs, user actions, and user values (e.g., a persona or user story). Methods with \textit{scenarios/context} as an input require a design situation or product scenario that the team has encountered, or a fictional situation that the user envisions. For example, Data Ethics Canvas \cite{dataethicscanvas} requires the user to formulate a scenario for which they need to plan data collection, storage, or opportunities for analysis. Methods with \textit{values} as an input require the user to formulate a list of personal, social, team, company, and/or project values to frame their decision making.
Methods with \textit{data} as input, for example Re-Shape \cite{reshape-website}, require the users to provide \textit{data} to teach data ethics and use the resulting data as a starting point for analysis and reflection. In a rare example, The Oracle for Transfeminist Technologies \cite{oracle} did not require any input to use the method, because the first step of the using the method involves filtering cards to create a problem space to generate futuristic concepts.

\textbf{\textit{Mechanics.}} We coded the action(s) expected from the users while using the method as its \textit{mechanic}. The mechanics we identified include: altering (n=10), creating (n=19), mapping (n=21), storytelling (n=23), filtering (n=17), and evaluating (n=23). Depending on the type of guidance provided by the method, there could be more than one type of mechanic for each method, hence these were non-exclusively coded. Methods with \textit{altering} as a mechanic expect users to edit a given worksheet as they follow the prescribed steps/guidelines in the method. For example, the Ethics Canvas \cite{ethicscanvas} provides a template for the users to collaboratively edit, move and add Post-It notes in appropriate sections to fill out the template. Methods using \textit{creating} as a mechanic encouraged a more conceptual and divergent approach whereby users produce artifacts through brainstorming, sketching, prototyping, and developing as they interact with the method, instead of providing existing artifacts for users to alter. Methods using \textit{mapping} expect users to draw connections or associations between method elements and artifacts created through the method. For example, Ethicography \cite{ethicography} is a method of value discovery that requires researchers to physically draw links that visualize the conversation change and growth through a design discussion; Value Source Analysis \cite{valuesourceanalysis} requires users to identify disagreements among stakeholders in order to conceptually map values for ``other environments.'' Methods engaging in \textit{storytelling} involve an act of role-playing, narrating stories, performing activities, or playing games as the users interact with the method. For example, in Judgement Call the Game \cite{judgementcallthegame}, users role-play as stakeholders and write a fictional review framed up by a combination of the rating card, stakeholder card, and ethical principle card that the player draws. Methods which require \textit{filtering} expect users to select scenarios, stakeholders, or draw cards, selecting salient options from a list of possibilities either provided by the method or created through the method. Methods with mechanics of \textit{evaluating} request users to assess the components provided by the method or artifacts produced through the method. For example, the Moral and Legal IT Deck \cite{moralitdeck} provides a wide range of critical questions, legal principles, and ethical principles for designers to follow and thereby evaluate ethically-related risks of the proposed new technology.

\textit{\textbf{Output.}} We coded the methods based on the tangible outcome that would be produced when using the method as the \textit{output}. The identified outputs include concepts (n=17), opportunities (n=23), evaluation results (n=16), values (n=20), users/stakeholders (n=4), user information (n=6), procedural information (n=4), and research outcomes (n=6). The majority of the kinds of outputs align with the list of inputs (listed above), although there is a clear change in their function of one being the input to use the method and the other being the result of the method usage. Depending on the method, it is possible that there is more than one possible or likely output for each methods, hence we non-exclusively coded for this descriptor. Methods with \textit{concepts} as an output result in new ideas, sketches, artifacts, and/or inspiration for future work. Methods resulting in \textit{opportunities} aid the user in locating ethical risks, recognizing future design possibilities through the method. For example, Ethical Disclaimer \cite{ethicaldisclaimer} is meant to allow the users to ``discuss for which of the unethical situations you will take responsibility.'' Methods with \textit{evaluation results} as outputs allow the user to assess their design through quantitative scores, ethical scores, design requirements, reflections, or other evaluation metrics. For instance, to illustrate the range of evaluation results, Ethical Design Scorecards \cite{ethicaldesignscorecards} provide an ``ethical score'' for the users to indicate the potential ethical valence of their design decision, and Re-Shape \cite{reshape-website} allows computer science students to evaluate their own decision making in the form of a reflection of their responsibilities towards data. Methods with \textit{values} as an output define a new mindset or outlook on what elements or abstract principles are most important in the users' design process. Methods with outputs such as \textit{users/stakeholders} and \textit{user information} have a similar definition as when they are used as inputs; however, as outputs the information about users is realized and produced through the methods. An output of \textit{procedural information} encompasses relevant and possible next steps and a future plan of action for decision making. Finally, methods with \textit{research outcomes} include artifacts produced using the method that are possible sources of future research or analysis, such as design research artifacts (e.g., user stories in speculative enactments \cite{speculative-enactments}) or co-design materials (e.g., in the Value Sensitive Action-Reflection Model \cite{actionreflectionmodel}).

\bigbreak
\textit{\textbf{Patterns of Input->Mechanics->Output.}} Based on our analysis, we have recorded the number of occurrences of each possible combination of input, mechanics, and output. As part of this approach, we mapped the interactions or patterns from ``input'' of the method as it was exclusively coded and present the most salient and frequently occurring interaction patterns. These interaction patterns aid us in describing the most typical ways in which the methods function as specification, and these patterns also elucidate possible opportunities for new or altered methods beyond these existing interaction patterns. We identified six common interaction patterns and provide their descriptions as follows:

\begin{itemize}
    \item \textit{Design Artifacts->Evaluating->Values/Evaluation Results}: This interaction pattern was found in methods that guide users to \textit{evaluate} existing \textit{design artifacts}, resulting in a range of \textit{evaluation results}. Such methods are intended to address existing product deficiencies, reveal ethical dilemmas of the system, and discover new values to be embedded in the design. Methods using this pattern include: Value dams and flows \cite{damsandflows}, Scalable assessments of information dimensions \cite{scalableassessments}, Moral and Legal Decks \cite{moralitdeck}, Inclusive Design Toolkit \cite{inclusivedesigntoolkit}, and GenderMag \cite{gendermag}.

    \item \textit{Design Artifacts->Mapping->Opportunities}: This interaction pattern was found frequently in methods that help users recombine, envision, and derive new design \textit{ opportunities} from \textit{existing artifacts} by \textit{mapping} out method elements or design space. Methods using this pattern include: Security Cards \cite{securitycards}, Multi-lifespan timeline \cite{multilifespand-timeline-co-design}, and the Inclusive Design Toolkit \cite{inclusivedesigntoolkit}.

    \item \textit{Design Artifacts->Storytelling->Values}: This interaction pattern is identified in methods which expect the user to interact with their design \textit{artifacts} through \textit{storytelling} or by playing games in order to explore, elicit, and engage with \textit{values} in designed artifacts.  Methods using this pattern include: Scalable assessments of information dimensions \cite{scalableassessments}, Ethics Canvas \cite{ethicscanvas}, Ethical Contract \cite{ethicalcontract}, and Envisioning cards \cite{envisioningcards}.

    \item \textit{Constraints->Creating->Concepts}: This interaction pattern is found in methods with design \textit{constraints} such as a design prompt, business timeline, or resource constraints, which results in \textit{creating} original or iterated \textit{concepts}. Methods using this pattern include: White Hat UX Patterns \cite{whitehat}, Value-oriented mock-up, prototype, or field deployment \cite{actionreflectionmodel}, Value Sensitive Action-Reflection Model \cite{actionreflectionmodel}, Moral Agent \cite{moralagent}, and Design with Intent \cite{designwithintent-book}.

    \item \textit{Problem Frame->Creating->Concepts}: This interaction pattern is used in methods that assist participants in \textit{generating} \textit{concepts} within a given or defined problem frame. Methods using this pattern include: Design for Social Accessibility Method Cards \cite{accessibilitymethodcards}.

    \item \textit{Scenario/Context->Creating->Concepts}: This interaction pattern is seen in methods which results in design concepts created within a defined, assumed or fictional scenario/context. Methods using this pattern include:  Value Sketch \cite{valuesketch}, Metaphor Cards \cite{metaphorscards-paper}, and Make It Critical \cite{makeitcritical}.

\end{itemize}

\bigbreak
\textbf{Existing frameworks/methods used.}
We sought to identify any existing design methods or frameworks that methods were built on, relied upon, or referenced. These frameworks are not translated into the method for the user, but rather they require the user to directly interact with these frameworks in order to successfully implement the ethics-focused method in their work. We identified two kinds of existing frameworks that were used: 1) established design methods; and 2) other standalone methods. Not all methods used existing frameworks, with only 25 out of the 63 methods representing this behavior. \textit{Established design methods} that were referenced (as listed in the Universal Methods of Design \cite{umod}) included: personas \cite{gendermag,adversarypersonas}, cognitive walkthrough \cite{empathycognitivewalkthrough}, stakeholder map \cite{stakholder-tokens}, scenarios \cite{value-scenarios,scenarioco-creationcards,workbooksprivacyfutures}, experience mapping \cite{makeitcritical}, cultural probes \cite{actionreflectionmodel}, qualitative research interviews (used as required technique in \cite{scenarioco-creationcards,value-orientedinterviews,scalableassessments}), and ethnography (used as a basic methodology for Ethnographically Informed \cite{ethnographically-informed}). These ethics-focused methods build upon or extend existing methods or approaches, facilitating the use of these methods with little prior preparation and expert knowledge. \textit{Other standalone methods} include less common methods, and frequently ethics-focused methods, with their own mechanics that are used in one of the methods we analyzed. Examples include: Ethics Canvas (another ethics focused method \cite{ethicscanvas}, which was used to inspire and build Data Ethics Canvas \cite{dataethicscanvas}); Design Heuristics (a card deck \cite{designheuristics} which was used in the cognitive walkthrough approach in the Idea Generation through Empathy method \cite{empathycognitivewalkthrough}); Ethical Disclaimer (another ethics-focused method which was used as an input in Ethical Contract \cite{ethicalcontract}); Linkography (used as a baseline framework \cite{linkography} in Ethicography \cite{ethicography}); and a combination of Value Scenarios \cite{value-scenarios}, Envisioning Cards \cite{envisioningcards}, and Value Sketch \cite{valuesketch} (used to build and follow steps in the Value Sensitive Action-Reflection Model \cite{actionreflectionmodel}). These methods require user(s)' existing knowledge in completing the method, including knowledge about the functioning of connecting methods or other related methods for using the ethics-focused method.

\bigbreak
\textbf{Design Process Implementation.}
We coded each method's suggested use or implementation across existing notions of design process stages (Table~\ref{designprocess}). We used an \textit{a priori} list of five design phases as suggested in Universal Methods of Design~\cite{umod} as an existing acknowledged mapping of design methods and process. We have chosen these five phases as they allow our analysis to build upon connections to implementation of the ethics-focused methods and the established mappings identified in \cite{umod}. Methods identified within \textit{Phase 1} included activities as planning, scoping, and definition, ``where project parameters are explored and defined'' (n=9). Methods in \textit{Phase 2} included activities such as ``exploration, synthesis, and design implications which are characterized by immersive research and design ethnography leading to design implications'' (n=40). Methods in \textit{Phase 3} included activities as ``concept generation and early prototype iteration, often involving generative and participatory design activities'' (n=27). Methods in \textit{Phase 4} included activities as ``evaluation, refinement, and production based on iterative testing and feedback'' (n=34). Finally, methods in \textit{Phase 5} included activities as ``launch[ing] and monitor[ing] the quality assurance testing of design to ensure readiness for market and public use, and ongoing review and analysis'' (n=7).
\small
\begin{table}
       \caption{Methods and Design Process Implementation}
    \label{designprocess}
    \begin{tabularx}{\textwidth}{p{.25\textwidth}X}
    \toprule
       \textbf{Design Process Phase \cite{umod}} &  \textbf{Methods}\\
    \midrule
\textbf{Phase 1} (planning, scoping, and definition) & Value Dams and Flows \cite{damsandflows}, Ethical Disclaimer \cite{ethicaldisclaimer}, Ethical Contract \cite{ethicalcontract}, Data Ethics Canvas \cite{dataethicscanvas}, 360 Review \cite{Zhou-360review}, Motivation Matrix \cite{Zhou-motivationmatrix},Inverted Model \cite{Zhou-invertedmodel}, Hippocratic Oath \cite{Zhou-hippocraticoath}, Design Fiction Memos \cite{DesignFictionMemos} \\
    \textbf{Phase 2} (exploration, synthesis, and design implications) & Value Value-Oriented Interviews\cite{whitehat,value-orientedinterviews}, Value Sensitive Action-Reflection Model\cite{actionreflectionmodel}, Value Scenarios \cite{value-scenarios}, Value Dams and Flows \cite{damsandflows}, Stakeholder Tokens \cite{stakholder-tokens}, Stakeholder Analysis \cite{stakeholderanalysis}, Speculative Enactments \cite{speculative-enactments}, Security Fictions \cite{securityfictions}, Security Cards \cite{securitycards},  Scenario Co-Creation Cards \cite{scenarioco-creationcards}, Scalable assessments of information dimensions \cite{scalableassessments}, Re-shape \cite{reshape-paper}, Normative Design Scheme \cite{normativedesignscheme},  Multi-lifespan Timeline \cite{multilifespand-timeline-co-design}, Moral and Legal Deck \cite{moralitdeck}, Moral Agent \cite{moralagent}, Model for Informed Consent \cite{modelinformedconsent}, Metaphor Cards \cite{metaphorscards-paper}, Make It Critical \cite{makeitcritical},HuValue \cite{huvalue}, Ethnographically informed inquiry on values and technology \cite{ethnographically-informed}, Ethicography \cite{ethicography}, Envisioning Cards \cite{envisioningcards}, Diverse Voices \cite{diversevoices}, Design with Intent \cite{designwithintent-book}, De-scription \cite{description}, Data Ethics Canvas \cite{dataethicscanvas}, Co-evolve technology and social structure \cite{co-evolvetechnology-social}, Blackmirror Brainstorming \cite{blackmirror-brainstorming}, Adversary Personas \cite{adversarypersonas}, Microsoft’s Inclusive Design Toolkit \cite{microsoftinclusive}, Layers of Effect \cite{Zhou-layersofeffect}, 360 Review \cite{Zhou-360review}, Motivation Matrix \cite{Zhou-motivationmatrix}, Inverted Model \cite{Zhou-invertedmodel}, Hippocratic Oath \cite{Zhou-hippocraticoath}, Timelines \cite{Timelines}, Design Fiction Memos \cite{DesignFictionMemos} \\
    \textbf{Phase 3} (concept generation and early prototype iteration) &
    White Hat UX Patterns \cite{whitehat}, Value Source Analysis \cite{valuesourceanalysis}, Value Sketch \cite{valuesketch}, Value Sensitive Action-Reflection Model \cite{actionreflectionmodel}, The Ethical Design Scorecards \cite{ethicaldesignscorecards}, Security Fictions \cite{securityfictions}, Security Cards \cite{securitycards}, Multi-lifespan Timeline \cite{multilifespand-timeline-co-design}, Moral Agent \cite{moralagent}, Metaphor Cards \cite{metaphorscards-paper}, Make It Critical \cite{makeitcritical}, Idea Generation through Empathy method \cite{empathycognitivewalkthrough}, HuValue \cite{huvalue}, Gender Mag \cite{gendermag}, Envisioning Cards \cite{envisioningcards}, Eliciting Values Reflections method \cite{workbooksprivacyfutures}, Design with Intent \cite{designwithintent-book}, Design for Social Accessibility Method Cards \cite{accessibilitymethodcards}, Data Ethics canvas \cite{dataethicscanvas}, Blackmirror Brainstorming \cite{blackmirror-brainstorming}, The Oracle for Transfeminist Technologies \cite{oracle}), Microsoft’s Inclusive Design Toolkit \cite{microsoftinclusive}, 360 Review \cite{Zhou-360review}, Dichotomy Mapping \cite{Zhou-dichotomymapping}, Inverted Model \cite{Zhou-invertedmodel}, Design Fiction Memos \cite{DesignFictionMemos} \\
    \textbf{Phase 4} (evaluation, refinement, and production) & White Hat UX Patterns \cite{whitehat}, Value Value-Oriented Interviews\cite{whitehat,value-orientedinterviews}, Value Source Analysis \cite{valuesourceanalysis}, Value Dams and Flows \cite{damsandflows}, Ethics and Inclusion Framework \cite{ethicsinclusionframework}, The Ethical Design Scorecards \cite{ethicaldesignscorecards}, Stakeholder Analysis \cite{stakeholderanalysis}, Speculative Enactments \cite{speculative-enactments}, Security Fictions \cite{securityfictions}, Scalable assessments of information dimensions \cite{scalableassessments}, Re-shape \cite{reshape-paper}, Normative Design Scheme \cite{normativedesignscheme},  Multi-lifespan Timeline \cite{multilifespand-timeline-co-design}, Moral Value Map \cite{moralvaluemap},  Moral and Legal Deck \cite{moralitdeck}, Making an ethical decision: A practical tool for thinking through tough choices \cite{markkulapracticaltool}, Judgement Call the Game \cite{judgementcallthegame}, Inclusive Design Toolkit \cite{inclusivedesigntoolkit}, Gender Mag \cite{gendermag}, Ethnographically informed inquiry on values and technology \cite{ethnographically-informed}, Ethics Canvas \cite{ethicscanvas}, Eliciting Values Reflections method \cite{workbooksprivacyfutures}, Design with Intent \cite{designwithintent-book}, The Tarot of Tech \cite{tarottechcards}, Spotify Design: Investigating Consequences with Our Ethics Assessment \cite{spotify}, Microsoft Inclusive Toolkit \cite{microsoftinclusive}, Layers of Effect \cite{Zhou-layersofeffect}, 360 Review \cite{Zhou-360review}, Dichotomy Mapping \cite{Zhou-dichotomymapping}, Maslow Mirrored \cite{Zhou-maslowmirrored}, Inverted Model \cite{Zhou-invertedmodel}, Monitoring Checklist \cite{Zhou-monitoringchecklist}, Design Fiction Memos \cite{DesignFictionMemos} \\
    \textbf{Phase 5} (launching and monitoring) & Speculative Enactments \cite{speculative-enactments}, Re-shape \cite{reshape-paper}, Gender Mag \cite{gendermag}, 360 Review \cite{Zhou-360review}, Inverted Model \cite{Zhou-invertedmodel}, Monitoring Checklist \cite{Zhou-monitoringchecklist}, Design Fiction Memos \cite{DesignFictionMemos} \\
    \bottomrule
    \end{tabularx}
\end{table}
\normalsize
Based on our analysis, we found the majority of the methods were designed for Phases 2, 3, and 4, with only rare examples in Phases 1 and 5. For methods coded as Phase 2, the focus was primarily on identifying design implications which aided the user in framing the problem space in more ethical ways; in contrast, methods coded as Phase 4 encouraged the user to build upon a generated design concept in more ethically-centered ways. As examples of Phase 2-focused methods, card decks were used in Adversary Personas \cite{adversarypersonas} to list potential adversaries in a particular design situation, while Envisioning Cards \cite{envisioningcards} were used to expand potential issues in the ``\textit{immediate context of use,}'' with the goal of envisioning the potential long-term impact of technology. The actions supported through these methods are likely to occur before concept generation, with the goal of framing the problem space by providing new ways of viewing the context. In Phase 3 and Phase 4, methods enable the production of ethically-focused designs and the evaluation of created or existing designs, respectively. For example, the Design for Social Accessibility Method Cards \cite{accessibilitymethodcards} provide users with ``concrete and real-life scenarios'' in Phase 3 to ``to generate accessible designs and appropriately engage deaf and hard-of-hearing users to incorporate social considerations.'' In Phase 4, the majority of methods provided ways for the user to evaluate their decisions or design outcomes, using guidance to refine their decisions. For example, the Ethics and Inclusion Framework~\cite{ethicsinclusionframework} aids the user in calculating the ``degree of inclusion of your product or service'' or facilitates ``assessment of potential negative outcomes'' for intended or unintended stakeholders.

\subsubsection{Articulation}
In this section, we describe the methods based on the way they are articulated to their respective audiences. We describe the ways methods are communicated to these audience(s) through two properties: 1) the type of guidance; and 2) the medium through which the method is communicated.

\bigbreak
\textbf{Type of guidance.}
We coded the descriptors that communicate the scaffolding or means of support to engage with the method in a way that is accessible to users as \textit{type of guidance} and the tangible form in which the guidance was provided as \textit{medium}. The type of guidance frames how the method is structured and conveyed to the user as instructional support or scaffolding \cite{Dennen2004-vo} in the following ways: steps (n=30), guidelines (n=23), framework (n=19), lens/perspectives (n=14), reflective questions (n=10), examples (n=22), heuristics (n=4), and case study (n=9). The method descriptions provided in the cited material frequently consisted of more than one kind of guidance, given the structure of the method, resulting in non-exclusive coding. If the method had multiple components or sub-components, we coded for all kinds of guidance provided, including any sub-structures of the method. \textit{Steps} are prescribed instructions to be followed by the user in order to interacting with the method in the provided order, whereas \textit{guidelines} do not insist on being followed in a specified order. For example, Diverse Voices \cite{diversevoices} provides ``main steps'' to be followed by the user starting with \textit{``Select a tech policy document''} and additional guidelines under each step to describe how and what kinds of tech policies documents to be selected. A \textit{framework} is a defined structure provided by the method developer in the form of a table, illustration, or schema. \textit{Lens/Perspectives} are possible attitudes or perspectives provided to focus the thought processes of the method user, while \textit{heuristics} are techniques that can be implemented non-deterministically in order to guide the user of the method. For example, White Hat UX Patterns~\cite{whitehat} lists a set of heuristics \textit{``to ensure ethical design''} outcomes, guided by heuristics such as: \textit{``Use data to improve the human experience,'}' and \textit{``Advertising without tracking.''} \textit{ Reflective questions} are posed as questions for the user to critically think through the ``input'' as intended by the ``core'' of the method.  \textit{Examples} are real world scenarios and/or visually illustrated guidance provided along with other types of guidance, while \textit{case studies} represent a real world design context through which the method is described rather than a standalone description of the method. Many of the methods proposed as part of the VSD methodology \cite{VSDbook} included case study-focused guidance that was represented as bound within a specific design decision, taking on characteristics of a \textit{case study}.

\bigbreak
\textbf{Medium.} We coded the descriptors that describe the tangible form of the methods as its \textit{primary medium}, which aided in communicating the above listed types of guidance to the user. The medium also inscribes how the method can be interacted with by the user in digital or physical format. The different medium types include: worksheets (n=17), templates (n=17), cards (n= 15), document/guidebook (n= 33), physical manipulatives (n=3), videos (n=1), and games (n=2). \textit{Worksheets} are documents where the user is asked to add specified information as they are working with the method, whereas a \textit{template} is a document that is expected to be used as a baseline reference in order to interact with using other components of the method. For example, HuValue \cite{huvalue} has a template with different value groups sectioned as a part of a circle which is intended to be used as a base with which to ``filter'' cards of user's choice under the value groups. Other physical media include cards or physical manipulatives, while digital media also include videos, and some methods could be presented in a combination of digital and physical forms. Design with Intent \cite{designwithintent-book} presents different ``lens'' through which design artifacts can be evaluated in the form of deck of physical color-coded \textit{cards}. A \textit{document/guidebook} could include a digital or physical standalone booklet that contains a method description and type of guidance for the user to refer as they are using the method during their design activity. Two methods---Moral Agent \cite{moralagent} and Judgement Call the Game \cite{judgementcallthegame}---were designed to encourage interactions in the form of a board game, consisting of a combination of several of the media described above. As another example of hybrid media, Moral Agent \cite{moralagent} consisted of a card deck to draw from, worksheets to write on, and two documents/guidebooks to filter values and describe the rules of the game.

\subsubsection{Language of Existing Methods}
We coded the descriptors that frame the language of these methods as \textit{sensitizing concepts,} drawing on a term by the same name that is often used to identify structure and conceptual foundations in grounded theory research \cite{Charmaz2008-nr}. These concepts provide a conceptual and methodological vocabulary that the method developers use to define the expected purpose or core of the method. Depending on the method focus, this vocabulary ranges from established social constructs (e.g., culture, gender); published policies (e.g., GDPR, EU Draft e-Privacy Regulation 2017); defined methodologies (e.g., critical design, VSD, speculative design, design fiction, co-design); known ethical or privacy concerns (e.g., user behavior change, cyber-security, data privacy, data ethics); defined human values (e.g., privacy, security, spirituality); commonly used interaction design concepts or methods (e.g., form, function, empathy, scenarios); and applied ethical concepts (e.g., justice, human rights, common good, utility). For example, GenderMag \cite{gendermag-teach,gendermag} relies upon the social category of ``gender'' as a sensitizing concept to frame the designer's construction of personas; this social construct shapes the method's purpose in engaging software developers in a more inclusive form of building technological artifacts and links the use of the method to broader social and academic conceptions of gender. The Moral and Legal Deck \cite{moralitdeck} cards used vocabulary from ``\textit{relevant rights, principles, definitions and responsibilities within the: EU General Data Protection Regulation 2016; EU Draft e-Privacy Regulation 2017; EU Network and Information Security Directive 2016; Cybercrime Convention 2001; and Attacks Against Information Systems Directive 2013}'' to design the content and guidelines provided through the method, thereby grounding design activity in legal definitions of privacy and data protection.

These sensitizing concepts had different functions in different methods, with some concepts being used across multiple methods in different ways to encourage or foreground specific mechanics, design judgments, or framings of design activity. For example, design fiction was defined as the means by which the designer should create ``design concepts'' in the Security Fictions method \cite{securityfictions}, whereas design fiction was used as an ideology with which the user could evaluate design concepts in Judgment Call the Game \cite{judgementcallthegame}. Design fiction was also treated as an opportunity space through which to explore and elicit values in relation to privacy in future technological artifacts in Eliciting Values Reflections method \cite{workbooksprivacyfutures}. Overall, we identified more than 80 sensitizing concepts across the set of methods we analyzed, and few sensitizing concepts appeared to be used consistently. Several key sensitizing concepts which did appear in multiple methods include: GDPR \cite{ethicaldesignscorecards,moralitdeck}, co-design \cite{actionreflectionmodel,speculative-enactments,valuesketch,multilifespand-timeline-co-design}, VSD (all methods within the VSD methodology), speculative design \cite{workbooksprivacyfutures,speculative-enactments,securityfictions,makeitcritical}, human values \cite{huvalue,ethicography,scenarioco-creationcards,actionreflectionmodel,valuesketch,valuesourceanalysis,value-orientedinterviews} and design fiction \cite{judgementcallthegame,speculative-enactments,securityfictions}. This analysis demonstrates that most methods include their own distinct vocabulary which is generally not shared or standardized across multiple methods,  illustrating both the variety in the existing ethics-focused methods and the lack of consistency across methods.

\section{Discussion}
In this paper, we have identified a range of descriptors of ethics-focused methods that allow us to identify existing mechanisms for ethical support, along with opportunities for the development, adaptation, and dissemination of new methods.
In this section, we return to the three stances towards design methods proposed by Gray~\cite{Gray2022-na} to characterize our contribution to the literature and describe opportunities for future work.
First, in Section~\ref{d1} we outline how the findings represent a codification-focused foundation for describing ethics-focused methods and point towards gaps in our current knowledge of ethics-focused methods, identifying spaces for new method development. Second, in Section \ref{d3} build upon our codification-focused analysis to identify how our analytic method vocabulary might be used to interrogate the performance and performativity of these methods in complex organizational contexts as an area of future work. Third, in Section \ref{definition} we conclude with a call for research and design work that engages ethics as a key dimension of design practice, pointing towards a new framing for scholarship that connects method design, method use, and method performance.

\subsection{Synthesis of Gaps in Current Knowledge of Ethics-Focused Methods for New Method Development} \label{d1}
We have identified a wide range of ethics-focused methods---many never before addressed in the academic literature, and none brought together as a collection previously beyond methodological ``families'' such as the VSD methods. This collection, in itself, represents a substantial contribution to HCI and design scholarship, allowing the collection to continue to grow, and providing a new analytic vocabulary to describe and frame both method knowledge in general and ethics-focused method knowledge in particular. When viewing our findings through a prescriptive- and presentation-oriented stance, we are able to use this preliminary landscape of ethics-focused methods to identify both areas of strong existing support and opportunities for new method development. First, we address the issues of access to these method materials due to their form of presentation and how this may point to future dissemination challenges. Second, we identify areas of strong and weak coverage of methods in relation to design process phases and describe opportunities for further method development.

A synthesis of the results from the content analysis facilitates additional focus on complexities that exist in terms of \textit{dissemination} of the methods' knowledge; \textit{codification} of these methods with respect to design process; and the \textit{performance} of these methods in everyday practice. While our analytic approach does not allow us to resolve questions or issues relating to dissemination challenges, we are able to observe a disconnect between the published formats of most of these method sources and their intended audience. According to our results on the published format, 44.4\% of the methods were published in the form of academic papers, and the majority of these methods were created for design practitioners or educators. However, this audience is frequently unable to access materials published in formal academic venues due to paywall restrictions. The lack of the access and awareness of these methods from the perspective of practitioners likely results in reduced adoption of the methods, even without accounting for other translational barriers observed by HCI researchers \cite{Gray2014-fk,Colusso2019-mw}. While our analytic approach does not allow us to resolve questions about dissemination challenges, our critique depends on the quantifiable results of comparing the method’s intended audience and published formats, which shows that most methods are behind a paywall. The distinction between academic (often paywalled) sources and practitioner sources is one delineation we observed in our analysis. Other presentation-oriented issues, such as the packaging of a method as a guide versus a physical game, could also impact dissemination and uptake in ways that could be investigated further in future work.

Additionally, as shown in Table~\ref{designprocess}, it is evident that the majority of methods we analyzed were research-focused (Phases 2 and 4), generatively-focused (Phase 3 and 4), or evaluation-focused (Phase 4). %This finding demonstrates the potential for exploring or creating methods for Phase 1 (project planning focused) and phase 5 (monitoring focused) work.
The creation of new methods focused on Phase 1 and 5 has the potential to be more impactful in creating a space for ethically-focused work, inscribing the problem space with ethical concerns (Phase 1) and continuously evaluating the work in a social context (Phase 5). One of the rare example of methods involved in Phase 1 of the process includes the Ethical Contract method \cite{ethicalcontract}, which enables a discussion among all the stakeholders on the project to clearly discuss and divide their ``ethical responsibilities'' prior to concept generation. This method, and others that might fit into this early stage of design work, shows potential in discussing, communicating, and formalizing ethical responsibilities across multiple stakeholders in everyday practice. Additionally, future work could also leverage the analytic vocabulary we have described,
and method developers and publishers could consider using Table~\ref{Codebook} as a framework to build, refine, or standardize their manuals.

\subsection{Methods as Supports for Ethical HCI Practice} \label{d3}
In the above section, we have presented opportunities to continue to grow the current landscape of prescriptive ethics-focused methods. Here, we extend our argument and present propositions about the nature of ethics-focused design methods that also consider performance in ecological setting, with a focus on identifying the potential \textit{ecological resonance} of these methods through a performance-oriented stance. While our analytic focus in this paper was on the knowledge contained within the method as it is codified by the method designer, our findings also point towards the potential performance of these methods in everyday practice, including inscriptions of underlying beliefs about practitioners, practice, and available resources. Methods we evaluated include mechanisms to aid in: addressing power dynamics and solving complexities due to organizational rules (e.g., Data Ethics Canvas \cite{dataethicscanvas}); bringing a balance between stakeholder requirements and designer intentions (e.g., A Value Sensitive Action-Reflection Model \cite{actionreflectionmodel}); facilitating realization of designers' ethical responsibilities and extending application of these responsibilities beyond instructional settings (e.g., Re-Shape \cite{reshape-paper});
guiding through self-provocation to evaluate the impacts of create technology (e.g., The Tarot of Tech \cite{tarottechcards}); bridging knowledge for practitioners from different disciplines (e.g., Idea Generation through Empathy method \cite{empathycognitivewalkthrough}); monitoring impacts of shipped products (e.g., Design Ethically-Monitoring Checklist \cite{Zhou-monitoringchecklist}); and providing ethical or critical knowledge, concepts, and vocabulary to be applied to support decision making (e.g. values through HuValue \cite{huvalue} and gender-inclusivity through GenderMag \cite{gendermag}).

Our synthesis also reveals several inscribed assumptions regarding the performance of these methods that could be used to support future research on ethics in HCI practice. First, the knowledge of these methods are presented using types of guidance that are intended to encourage certain patterns of performance. For instance, heuristics are used in the White Hat UX patterns \cite{whitehat} method, with the underlying assumption that these heuristics can be applied without the impedance of existing business forces or other forms of complexity beyond the designer themself. Second, the methods reveal beliefs that designers already have the vocabulary and capacity to express their social responsibility and have the capability to take responsibility for positive social impact (e.g., as required to list in Ethical Disclaimer \cite{ethicaldisclaimer}). Third, the methods reveal beliefs that practitioners are able to evaluate their decisions based on concepts such as utility, human rights, and justice, which often resist quantification. Using these inscribed assumptions as a point of departure, scholars may focus future research to investigate the degree to which these inscriptions impact performance, how these inscriptions impact the resonance of methods in particular practice contexts, and how different forms of ethics operationalization in methods link to encountering or impacting specific kinds of complexity in practice.

These qualities of the intended performance of methods' knowledge in practice also brings to the foreground the resonance of these methods with the constraints and complexity of practice settings, and the capacity and existing knowledge of the practitioners. Building on Stolterman's \cite{Stolterman2008-ho} concept of \textit{rationality resonance}, which highlights the relationship between \textit{suggested} vs. \textit{existing} practice, we can further question the barriers to adoption of these methods in practice settings, identifying spaces where existing assumed knowledge of practitioners is incomplete; spaces where the agency and power of designers is not available to the extent that methods might assume; and spaces where ethical complexity across multiple stakeholder positions is unaccounted for \cite{Gray2019-wa}. While we cannot resolve this issue of performance in relation to the methods we analyzed in the context of this paper, we do propose that future work could address the role of methods as an emergent ``new rationality'' that could promote ethical practices, while also guarding against method descriptions and codification that lack resonance with the ethical design complexity present in everyday work practices.

\subsection{Creating a Space for Scholarship and Design of Ethics-Focused Methods} \label{definition}
Prior design theory literature defines methods as a source of design knowledge that enables or supports design activity, acting as a toolset to support the designer's judgment and action throughout their design work. Based on the sensitizing concepts and core of these methods, we have identified a range of underlying assumptions regarding the use of these methods to discover knowledge, identify new possibilities, and locate hidden assumptions. These attitudes or stances towards engagement with knowledge range from uncovering ethical components of a design situation or problem through \textit{in situ} speculation (e.g., Speculative Enactments \cite{speculative-enactments}); considering ethical evaluation through acts of iterating and futuring (e.g., The Oracle for Transfeminist Technologies \cite{oracle}); fostering innovation through unique value propositions by uncovering neglected or negatively impacted user groups in a use scenario (e.g., The Ethics and Inclusion Framework \cite{ethicsinclusionframework}); influencing designers' thinking to design for sustainable and non-deceptive behavior change (e.g., Design with Intent \cite{designwithintent-book}); helping to foreground and map the underlying intentions and world-view of a designer (e.g., Description \cite{description}); and providing a means for designers to operationalize ethics in their design process.

The primary aim of the knowledge contained in the methods we have analyzed in this paper is ethical impact---influenced by ethical theories, values, and frameworks---represented through the \textit{core} of the methods. We chose to define the collection as containing ``ethics-focused methods'' to differentiate these methods from conventional design methods, conceptual frames, or theoretical commitments. In this way, the resulting collection of 63 ethics-focused methods include prescriptive forms that are actionable and potentially performative on the part of designers. Thus, the function of the method revealed through this \textit{embedded knowledge} allows designers to convert ethics-focused discovery into design outcomes. We have identified several means by which this translation might occur, including: converting a prescribed value into a concern by describing how it is present in the design context (e.g., Moral Value Map \cite{moralvaluemap}); maximizing the ethical valence of a design situation by considering many ethical theories together (e.g., Normative Design Scheme \cite{normativedesignscheme}); bringing legal, moral and ethical policies and values together (e.g., Moral and Legal IT Deck \cite{moralitdeck}); introducing user-centric concepts into different disciplines (e.g., Idea Generation through Empathy method \cite{empathycognitivewalkthrough}); broadening the scope of human values such as privacy (e.g., Privacy Futures through Design Workbooks \cite{workbooksprivacyfutures}); quantifying ethical decisions for users to improve their decision making (e.g., The Ethical Design Scorecards \cite{ethicaldesignscorecards}); and introducing critical and feminist constructs such as gender to expand the horizons of technology design (e.g., Gender Mag \cite{gendermag}). This embedded knowledge aids the designer in translating complex ethical concepts into normatively-informed work practices, bridging the liminal space between awareness and action.

Our identification of both a collection of ethics-focused methods and an analytic vocabulary to describe elements of these design methods points towards a new space for research and design practices that can better describe and support ethical dimensions of HCI work. First, this work serves as the beginning point for the intentional collection and curation of design methods with an ethical focus. We call on scholars and practitioners to add to this collection through the addition of formal methods, further description of adaptation and performance of these methods in specific practice contexts, and identification of new types of ethical inscription beyond the method cores we have identified in our analysis. Second, this work provides a new language to describe knowledge contained within methods, building on the work of Gray~\cite{Gray2022-na} and others. This vocabulary has uptakes not only for building and extending a collection of ethics-focused methods, but also in providing more analytic precision in describing how methods intersect with practice in a performative stance. Third, this work lays the foundation for both the creation of new methods and the identification of strategies to better inscribe ethical impact into methods. Future work can investigate and evaluate ideal spaces for new methodological support and compare differing combinations of cores, presentation formats, and/or mechanics for ethically-engaged design.

\section{Implications and Future Work}
Our findings include the identification of various descriptors embedded in existing ethics-focused methods, and the discussion reveals even more complexities and underlying assumptions about the practical use of these methods that relate to their specification, dissemination, and performance. Building on this work, there is substantial potential for future research through investigation of issues relating to method adaption, evolution, and the resonance of these methods with everyday work practices. Work in this area may lead to the identification of productive areas for the creation of future ethics-focused methods, while also pointing towards barriers to adoption of existing methods in practitioner discourses and work practices.

While we have identified a large set of ethics-focused methods, we do not claim that our collection is decidedly complete. Due to the wide range of codification approaches and dissemination strategies used in the set of methods we were able to identify, collect, and analyze, we anticipate that there are likely other existing and emergent sources that could enhance our collection. Thus, we do not rest the implications and contribution of this work on this initial collection being objectively ``complete''; rather, our primary contribution is focused on this collection as a foundation and the ways we have analyzed and attached descriptors to this collection of methods, revealing opportunities for new methods, refinement of existing methods, and means of standardizing language among methods to increase portability and adaptation in everyday design work.

Further investigation into the practical use and popular awareness of these methods may provide opportunities for method developers to consider alternative dissemination strategies. Future work may productively focus on studying the creation of these methods, revealing the considerations  method developers take into account, and the ways in which they constrain or include aspects of ethical design complexity into the methods they create.

Building on our findings, we are able to identify opportunities for the creation of new ethics-focused methods, and additional practices that may result in more resonant forms of methods dissemination, design process implementation, iteration and adaptation, and translational opportunities among design practitioners, educators, and researchers. First, we underscore the need to disseminate and distribute methods to the intended audience in more accessible and public formats, including the potential creation of channels to sharing material between researcher and practitioner communities. Second, we describe a substantial gap in the provision of design methods that support phases 1 and 5, including ethical engagement with the framing of a problem space by considering ethical responsibilities across all stakeholders (Phase 1) and iteratively evaluating a product in a social context after launching in the market (Phase 5). The creation of new methods to address these spaces, or the intentional adaptation of existing methods to support these forms of inquiry could bring substantial value to design conversations in these areas of practice. Third, methods could be further evaluated in terms of their fit and portability, encouraging increased iterative use of a range of ethics-focused methods across a range of design activities. The input->mechanic->output patterns reveal rich opportunities for this interplay among methods, yet the instructions and media of the methods we have evaluated show distinct differences in approach that make this ad hoc assemblages of methods difficult or unlikely under the pressures of everyday practice. Fourth, we identify potential  opportunities to build prescriptive methods that rely upon theoretical commitments (i.e., Table~\ref{types}), activating those concepts through new ethics-focused methods to support design practice. Finally, we propose the creation or articulation of value-focused design frameworks for a combination of stakeholders to build ethical alignment and engage members of a multi-disciplinary team, bringing resonance across multiple practice contexts. Successfully engaging with these implications and provocations for future work may substantively impact the ethical awareness of design and technology practitioners, while also pointing towards needs and gaps in design education practices, where a core set of methods is often learned.

\section{Conclusion}
In this paper, we present a content analysis of 63 ethics-focused methods intended for use in design and technology practice. We map the current landscape of ethical support and tools by characterizing the collection of these methods along multiple dimensions, including the ways in which they operationalize ethics, their intended primary audience and context of use, the core of the methods, their interaction qualities, and the ways in which these methods are described. We provide these methods as an initial collection, alongside a set of descriptors that mark the existing landscape of ethical support for researchers, practitioners, and educators. We propose a definition for ethics-focused methods and identify means of making these methods more resonant with HCI and design practice, articulating multiple areas of future work to support method development and design practice.

\begin{acks}
This work is funded in part by the National Science Foundation under Grant No. 1909714.
\end{acks}

\bibliographystyle{ACM-Reference-Format}
\bibliography{methods}

%\appendix

\end{document}